\documentclass[11pt]{article}
\usepackage{hyperref}
\pdfoutput=1
\begin{document}
\title{Unstable Leidenfrost Drops on Roughened Surfaces}
\author{Jonathan B. Boreyko and Chuan-Hua Chen \\
\\\vspace{6pt} Department of Mechanical Engineering and Materials Science, 
\\ Duke University, Durham, NC 27708, USA}
\maketitle
\begin{abstract}
Drops placed on a surface with a temperature above the Leidenfrost point float atop an evaporative vapor layer.
In this fluid dynamics video, it is shown that for roughened surfaces the Leidenfrost point
depends on the drop size, which runs contrary to previous claims of size independence.  The thickness of the vapor layer is known to increase with
drop radius, suggesting that the surface roughness will not be able to penetrate the vapor layer 
for drops above a critical size.  This size dependence was experimentally verified: at a given roughness and temperature, 
drops beneath a critical size exhibited transition boiling
while drops above the critical size were in the Leidenfrost regime.  These Leidenfrost drops were unstable; upon evaporation down to the 
critical size the vapor film suddenly collapsed.

\end{abstract}
\section{Introduction}
The video includes all necessary information and is available in \href{}{large} and \href{}{small} sizes.

\end{document}